\documentclass[prd,aps,showpacs,superscriptaddress]{revtex4}
\usepackage{graphicx}
\usepackage{bm}
\usepackage{dcolumn}
\def\ds{\displaystyle}

\def\pr{\prime}
\def\Dot{\!\cdot\!}
\def\al{\alpha}
\def\be{\beta}
\def\ga{\gamma}
\def\de{\delta}
\def\la{\lambda}
\def\ro{\rho}
\def\ep{\varepsilon}
\def\arcsinh{\mbox{\rm arcsinh}}

\def\erf{\mbox{\rm erf}}
\begin{document}
\title{Measuring the neutrino mass using intense photon and neutrino beams}
\author{Duane A. Dicus}
\affiliation{Center for Particle Physics and
Department of Physics, University of Texas, Austin, Texas 78712}
\author{Wayne W. Repko}
\affiliation{Department of Physics and Astronomy, Michigan State
University, East Lansing, Michigan 48824}
\author{Roberto Vega}
\affiliation{Department of Physics, Southern Methodist University, Dallas, Texas 75275}
\date{\today}

\begin{abstract}
We compute the cross section for neutrino-photon scattering taking into
account a neutrino mass.  We explore the possibility of using intense
neutrino beams, such as those available at proposed muon colliders,
together with high powered lasers to probe the neutrino mass in
photon-neutrino collisions.
\end{abstract}
 \maketitle

\section{Introduction}

Several experiments studying solar, atmospheric and reactor neutrinos
accumulated over the past several years provide an increasing body of
evidence supporting the existence of neutrino oscillations
\cite{conrad,SLAC}. The existence of neutrino oscillations will require a
significant departure from the the Standard Model. Oscillations imply that
at least one of the neutrinos is massive and that lepton number is not
conserved. The oscillation probability depends on the mass difference
between the oscillating neutrinos and is insensitive to the value of the
neutrino mass.  The possible values of the mass difference are small,
lying typically in the range $10^{-5}$ to $10^{-3}$ eV.  The experimental
limits on the muon and tau neutrino masses come from kinematic
distributions in weak decays.  They are \cite{pdg} $m_{\nu_{\mu}}<0.17$\,
MeV and $m_{\nu_{\tau}}<18.2$\, MeV.  If oscillations do occur, the mass
of $\nu_{\tau}$ is expected to be within less than 1 eV from that of
$\nu_{\mu}$, and, therefore, the determination of the muon neutrino mass
is of special importance. The experimental measurement setting a limit on
the $\nu_{\mu}$ mass is based on the kinematics of pion decay at rest and
is dominated by difficult to control systematic effects.  Consequently, it
is of special particular interest to explore all possible other processes
that may be sensitive to the neutrino mass.

We note that neutrino photon scattering, with and without photon production in
both the non-relativistic and relativistic regimes, has been studied
extensively within the standard model \cite{DR}. Astrophysical implications
have been considered in, for example, \cite{teplitz} and the effect of
scattering in a background magnetic field has also been investigated
\cite{sha,chyi}.

This paper attempts to exploit the fact that these cross sections vary as
the square of the neutrino mass to determine whether it might be possible
to measure the elastic scattering cross section for values of the muon
neutrino mass below its current limit cited above.  We find that it is not
possible with neutrinos from a facility on the order of the muon collider
under current discussion.  We provide an indication of the kind of
facility that would support such a measurement.  In Section 2 we derive
the needed formulae and evaluate the cross section and in Section 3 we
sketch the scope of an experiment to measure $m_{\nu}$. This is followed
by a discussion.

\section{Contributions to the  $\bm{\ga\nu\to\ga\nu}$ cross section}
\subsection{ ${\bm Z^0}$ exchange}

\begin{figure}[h]
\centering\includegraphics[height=1.5in,clip]{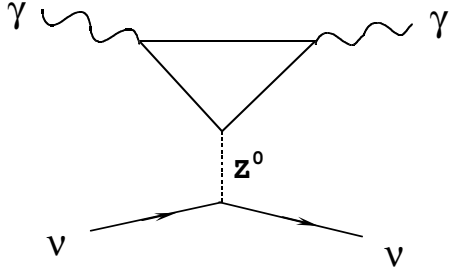}
\caption{\footnotesize A typical fermion triangle diagram for the
$\ga\nu\protect\to\ga\nu$ amplitude is shown. The particle in the loop can
be an electron, up quark, down quark, etc.\label{ztri}}
\end{figure}

A typical contribution to the amplitude for photon-neutrino
scattering due to a fermion loop with $Z^0$ exchange is
illustrated in Fig.\,\ref{ztri}. Because of Yang's theorem
\cite{Yang}, this amplitude will contain a factor of the neutrino
mass $m_{\nu}$\,\cite{GellMann}. For center of mass energies low
compared to the $Z^0$ mass,
the effective neutral current coupling between the charged
fermions and neutrinos due to $Z^0$ exchange is
\begin{equation} \label{lz}
{\cal L}_Z =
\frac{G_F}{\sqrt{2}}\bar{\nu}(x)\ga_{\mu}(1+\ga_5)\nu(x)\bar{f}(x)\ga_{\mu}
\left((t_3(f)-2q_f\sin^2\theta_W)+t_3(f)\ga_5\right)f(x)\,,
\end{equation}
where $t_3(f)$ is the third component of the fermion weak isospin,
and $q_f$ is the fermion charge in units of the proton charge.
When the coupling, Eq.\,(\ref{lz}) is combined with the fermion
electromagnetic coupling, only the weak axial vector contribution
survives. For a particular fermion, the triangle amplitude takes
the form
\begin{equation}
{\cal
A}_Z=\frac{G_F}{\sqrt{2}}\bar{u}(p_2)\ga_{\mu}(1+\ga_5)u(p_1)t_3(f){\cal
M}_{\la\ro,\mu}\ep_{\la}(k_1)\ep_{\ro}^*(k_2)\,,
\end{equation}
with
\begin{equation} \label{zamp}
{\cal M}_{\la\ro,\mu}=\frac{4\al q_f^2C_f}{\pi t}\left[\frac{1}{2}
+ \frac{m_f^2}{t}\int_0^1\frac{dx}{x}\ln\left(1 -
\frac{t}{m_f^2}x(1 -
x)\right)\right]\ep_{\la\ro\al\be}k_{1\,\al}k_{2\,\be}(k_1-k_2)_{\mu}\,,
\end{equation}
where $C_f$ is the fermion color factor and $t = 2k_1\Dot k_2$ is
the invariant momentum transfer. The remaining integral can be
evaluated as
\begin{equation}\label{arcsinh}
\int_0^1\frac{dx}{x}\ln\left(1 + yx(1-x)\right) =
2\arcsinh^2\!\left(\sqrt{\frac{\ds y}{\ds 4}}\,\right)\,,
\end{equation}
when $y\geq 0$. From Eq.\,(\ref{arcsinh}), ${\cal M}_{\la\ro,\mu}$
is expressible as
\begin{equation}
{\cal M}_{\la\ro,\mu} = \frac{4\al q_f^2C_f}{\pi
t}\left[\frac{1}{2}+2\frac{m_f^2}{t}\arcsinh^2\!\left(\sqrt{\frac{\ds
-t}{\ds
4m_f^2}}\,\right)\right]\ep_{\la\ro\al\be}k_{1\,\al}k_{2\,\be}(k_1-k_2)_{\mu}\,.
\end{equation}

With these simplifications, the amplitude for elastic scattering
can be written
\begin{eqnarray}
{\cal A}_Z & = & \frac{G_F}{\sqrt{2}}\frac{4\al}{\pi
t}\bar{u}(p_2)\ga_{\mu}(1+\ga_5)
u(p_1)\ep_{\la\ro\al\be}k_{1\,\al}k_{2\,\be}\ep_{\la}(k_1)\ep_{\ro}^*(k_2)(k_1-k_2)_{\mu}
\nonumber \\ [4pt] &
&\times\sum_f\left\{t_3(f)q_f^2C_f\left[\frac{1}{2}+2\frac{m_f^2}{t}\arcsinh^2\!
\left(\sqrt{\frac{\ds -t}{\ds 4m_f^2}}\right)\right]\right\}\,.
\end{eqnarray}
Using the equations of motion, we have
\begin{eqnarray}
(k_1-k_2)_{\mu}\bar{u}(p_2)\ga_{\mu}(1+\ga_5)u(p_1) & = &
(p_2-p_1)_{\mu}\bar{u}(p_2)\ga_{\mu}(1+\ga_5)u(p_1) \nonumber \\
[4pt] & = & 2im_{\nu}\bar{u}(p_2)\ga_5u(p_1)\,
\end{eqnarray}
and
\begin{equation}
{\cal A}_Z  = \frac{8iG_Fm_{\nu}\al}{\sqrt{2}\pi
t}\bar{u}(p_2)\ga_5u(p_1)\ep_{\la\ro\al\be}k_{1\,\al}k_{2\,\be}\ep_{\la}(k_1)\ep_{\ro}^*(k_2){\cal
B}(t)\,,
\end{equation}
with
\begin{equation} \label{B}
{\cal B}(t) =
\sum_f\left\{t_3(f)q_f^2C_f\left[\frac{1}{2}+2\frac{m_f^2}{t}\arcsinh^2\!\left(\sqrt{\frac{\ds
-t}{\ds 4m_f^2}}\,\right)\right]\right\}\,.
\end{equation}
In Eq.\,(\ref{B}), the sum $\sum_ft_3(f)q_f^2C_f$ vanishes for any
generation, providing the anomaly cancellation, and we effectively have
\begin{eqnarray}
{\cal B}(t) & \to &\frac{2}{t}\sum_ft_3(f)q_f^2C_fm_f^2
\arcsinh^2\!\left(\sqrt{\frac{\ds -t}{\ds 4m_f^2}}\,\right)
\nonumber \\ [4pt] & = &\frac{2}{t}{\cal C}(t)\,.
\end{eqnarray}
The squared amplitude then has the form
\begin{equation}
\left|{\cal A}_Z\right|^2 =
\frac{16^2G_F^2m_{\nu}^2\al^2}{2\pi^2t^4}|{\cal C}(t)|^2
|\bar{u}(p_2)\ga_5u(p_1)\ep_{\la\ro\al\be}k_{1\,\al}k_{2\,\be}\ep_{\la}(k_1)\ep_{\ro}^*(k_2)|^2\,,
\end{equation}
which, with the spin sum
\begin{equation}
\sum_{\rm
spin}|\bar{u}(p_2)\ga_5u(p_1)\ep_{\la\ro\al\be}k_{1\,\al}k_{2\,\be}\ep_{\la}(k_1)\ep_{\ro}^*(k_2)|^2
= -t^3\,,
\end{equation}
yields
\begin{equation}
\sum_{\rm spin}\left|{\cal A}_Z\right|^2 =
\frac{16^2G_F^2m_{\nu}^2\al^2}{2\pi^2(-t)}|{\cal C}(t)|^2\,.
\end{equation}
Using the expansion $\arcsinh^2(x) = x^2 - x^4/3$, ${\cal C}(t)$
can be expanded to the lowest order in $t/m_f^2$ as
\begin{equation} \label{c(t)1}
{\cal C}(t) = -\frac{t^2}{48}\sum_f\frac{t_3(f)q_f^2C_f}{m_f^2}\,.
\end{equation}
\begin{figure}[t]
\centering\includegraphics[height=2.4in]{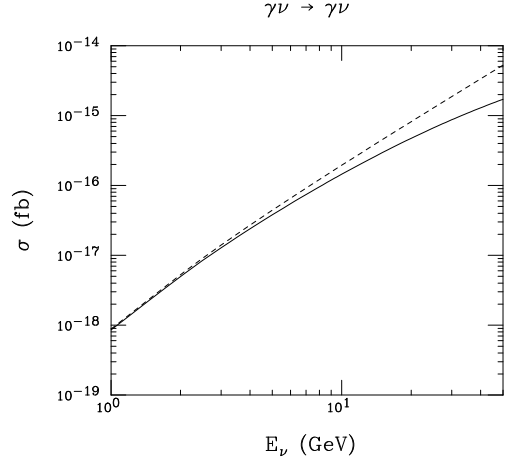}
\caption{\footnotesize The exact (solid) and leading order (dashed) cross
sections for $\gamma\nu\protect\to\gamma\nu$ are shown for a range of neutrino
energies $E_{\nu}$. The neutrino mass is 100 keV, the photon energy is 10 eV,
and the up and down quark masses are $m_u = 3\,$MeV and $m_d = 6\,$MeV.
\label{sig}}
\end{figure}

The differential cross section may now be calculated using
\begin{eqnarray}
\frac{d\sigma}{dt} & =
&\frac{1}{16\pi(s-m_{\nu})^2}\frac{1}{2}\sum_{\rm spin}\left|{\cal
A}_Z\right|^2 \nonumber \\ [4pt] & =
&\frac{16^2G_F^2m_{\nu}^2\al^2}{(4\pi)^3(s-m_{\nu}^2)^2}\frac{|{\cal
C}(t)|^2}{(-t)}\,,
\end{eqnarray}
where $-(s - m_{\nu}^2)^2/s\leq -t\leq 0 $, and the total cross
section is
\begin{equation} \label{sigtot}
\sigma =
\frac{16^2G_F^2m_{\nu}^2\al^2}{(4\pi)^3(s-m_{\nu}^2)^2}\int_0^{(s-m_{\nu}^2)^2/s}\frac{dt}{t}|{\cal
C}(-t)|^2\,.
\end{equation}
To the leading order, Eq.\,(\ref{c(t)1}) can be used to obtain
\begin{equation}
\int_0^{(s-m_{\nu}^2)^2/s}\frac{dt}{t}|{\cal C}(-t)|^2 =
\frac{1}{16^2}\frac{1}{36}\left(\sum_f\frac{t_3(f)q_f^2C_f}{m_f^2}\right)^2\frac{(s-m_{\nu}^2)^8}{s^4}\,.
\end{equation}
This results in the leading order cross section
\begin{equation} \label{siglead}
\sigma =
\frac{G_F^2m_{\nu}^2\al^2}{144(4\pi)^3}\left(1+\frac{m_e^2}{3m_d^2}-\frac{4m_e^2}{3m_u^2}\right)^2\frac{(s-m_{\nu}^2)^6}{m_e^4s^4}\,.
\end{equation}
Note the $m_e^{-4}$ dependence of Eq.\,(\ref{siglead}).

The exact result for $\sigma$ can be obtained by numerical integration, and is
shown in Fig.\,\ref{sig} together with the leading order result,
Eq.\,(\ref{siglead}). It is clear that the leading order cross section is an
overestimate. A few specific numbers are listed in Table\,\ref{tab:1}.

\begin{figure}[t]
\centering\includegraphics[height=2.4in]{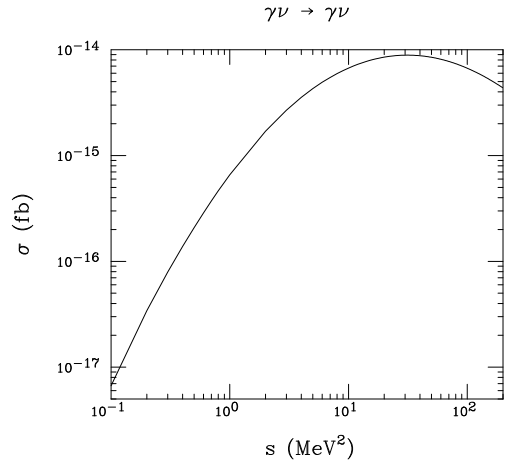}
\caption{The cross section $\sigma(\gamma\nu\protect\to\gamma\nu)$ is
shown for a range of center of mass energy squared. The maximum occurs at
$s=31.4$\,MeV$^2$. \label{sdep}}
\end{figure}

\begin{figure}[t]
\hfill\includegraphics[height=2.4in]{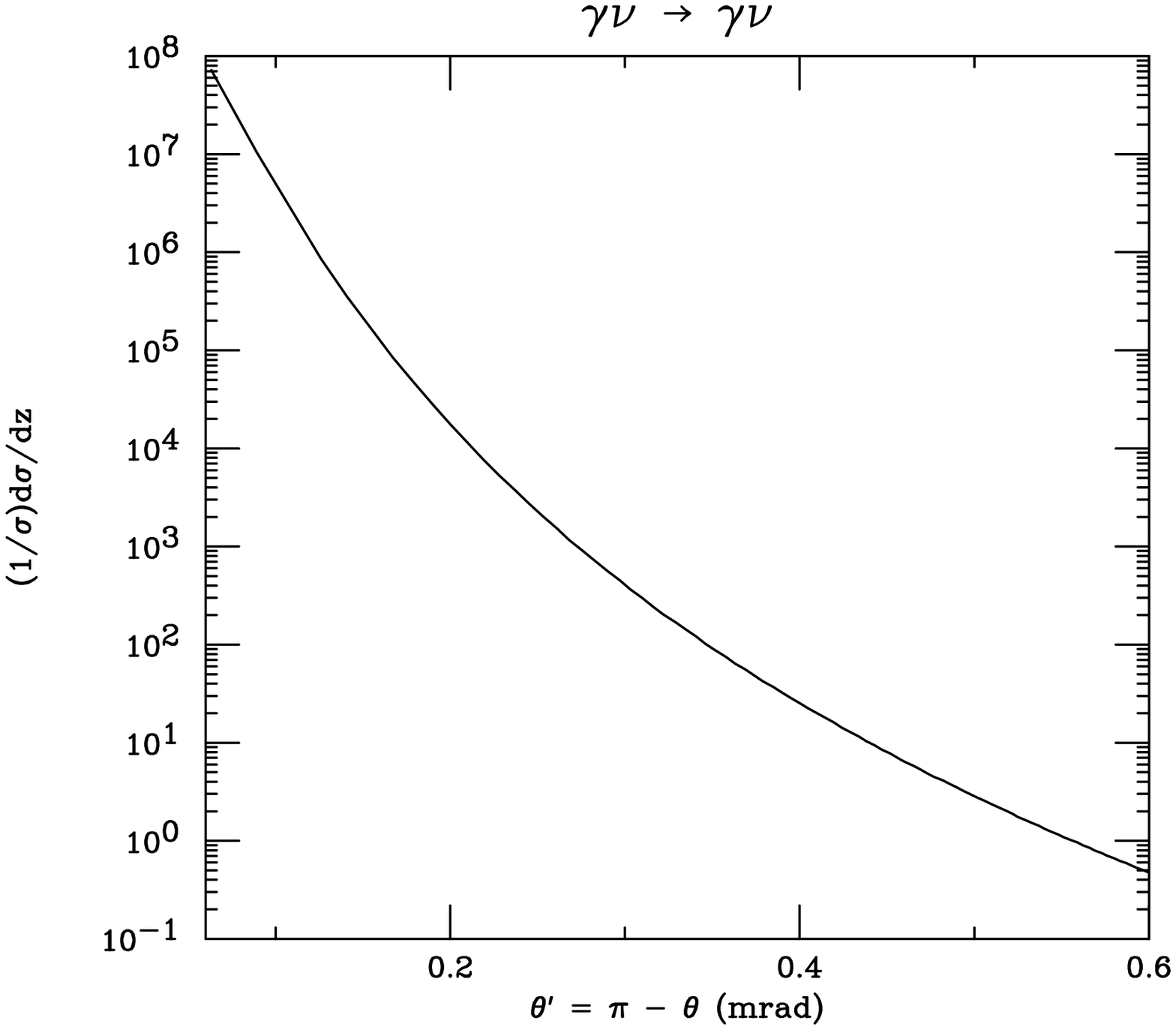}%
\hfill\includegraphics[height=2.4in]{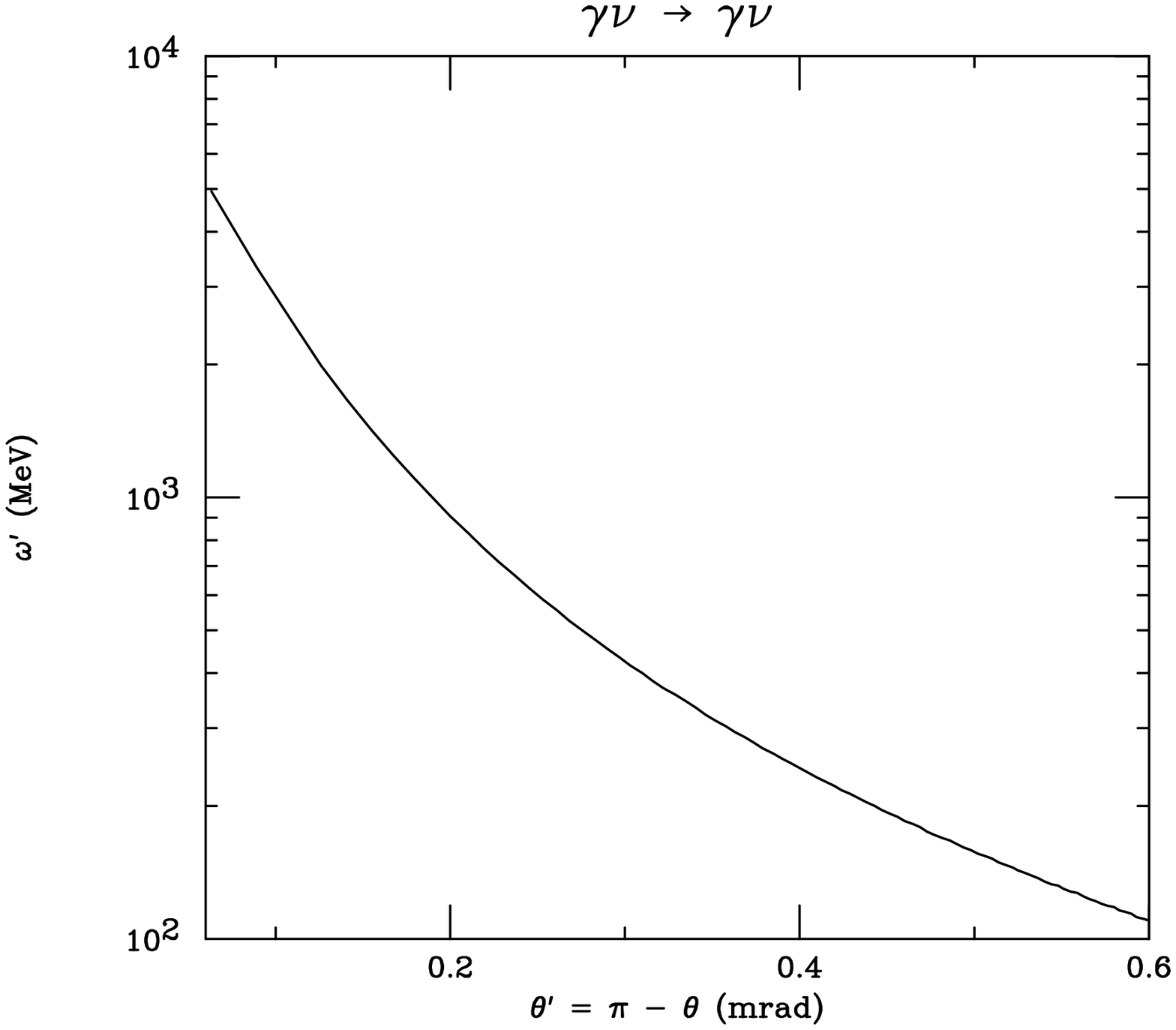}\hspace*{0pt\hfill}
\caption{In the left panel, backward peak in the photon angular
distribution is shown as a function of the angle $\theta^{\prime}$ away
from the backward direction. The right panel shows the energy
$\omega^{\prime}$ of the backward scattered photon in MeV for a neutrino
beam with $E_{\nu}=10$\,GeV.\label{dist}}
\end{figure}

The dependence of the cross section on the square of the center of mass
energy $s$ can be seen in Fig.\,\ref{sdep}. The maximum value is $\sigma =
8.89\times 10^{-54}\,$ cm$^2$, which corresponds to $s = 31.4\,$ MeV$^2$.
In the laboratory frame, where the invariant momentum transfer $t$
is
\begin{equation}
t =
-\frac{2\omega^2(E_{\nu}+p_{\nu})(1-\cos\theta)}{\omega+E_{\nu}-(\omega-p_{\nu})\cos\theta}\,,
\end{equation}
the angular distribution is extremely sharply peaked in the
backward direction, as shown in left panel of Fig.\,\ref{dist}.
The energy of the backward scattered photon is shown in the right
panel of Fig.\,\ref{dist}, and it, too, is sharply peaked.

\subsection{Higgs exchange}

In addition to contributions from $Z^0$ exchange, the standard
model fermion-Higgs-boson coupling, $-gm_f\bar{f}fH^0/2m_W$, gives
rise to a scalar triangle diagram similar to Fig.\,\ref{ztri},
with $H^0$ replacing $Z^0$. The amplitude associated with this
contribution has the form
\begin{equation} \label{hamp}
{\cal
A}_H=\frac{G_Fm_{\nu}}{\ds\sqrt{2}}\frac{m_f^2}{m_H^2}\bar{u}(p_2)u(p_1){\cal
M}_{\mu\nu}\ep_{\mu}(k_1)\ep_{\nu}^*(k_2)\,,
\end{equation}
with
\begin{equation} \label{scalar}
{\cal M}_{\mu\nu}=\frac{4\al
q_f^2C_f}{\pi}\int_0^1dx\int_0^{(1-x)}dy\frac{(1-4xy)}{(m_f^2-txy)}\,(k_1\Dot
k_2\de_{\mu\nu}-k_{2\mu}k_{1\nu})\,.
\end{equation}
Because of the spinor factors in Eqs.\,(\ref{zamp}) and
(\ref{hamp}), there is no interference between the $Z^0$ and $H^0$
amplitudes. We can, therefore, assess the importance of the scalar
amplitude by simply calculating corresponding cross section. Using
the result
\begin{equation}
\int_0^1dx\int_0^{(1-x)}dy\frac{(1-4xy)}{(m_f^2-txy)}=\frac{2}{t}\left[1+\frac{(4m_f^2-t)}{t}\arcsinh^2\!\left(\sqrt{\frac{\ds
-t}{\ds 4m_f^2}}\,\right)\right]\,,
\end{equation}
the amplitude for a particular fermion takes the form
\begin{equation}
{\cal A}_H=\frac{8G_Fm_{\nu}\al q_f^2C_f}{\sqrt{2}\pi
t}\frac{m_f^2}{m_H^2}\bar{u}(p_2)u(p_1)(k_1\Dot
k_2\ep_1\Dot\ep_2^*-k_2\Dot\ep_1 k_1\Dot\ep_2^*){\cal
B}^{\pr}(t)\,,
\end{equation}
with
\begin{equation}
{\cal
B}^{\pr}(t)=\left[1+\frac{(4m_f^2-t)}{t}\arcsinh^2\!\left(\sqrt{\frac{\ds
-t}{\ds 4m_f^2}}\,\right)\right]\,.
\end{equation}
In this case, the spin averaged squared matrix element is
\begin{equation}
\frac{1}{2}\sum_{\mbox{\rm spin}}|{\cal A}_H|^2 =
\frac{16G_F^2m_{\nu}^2\al^2}{\pi^2}\left(\frac{m_f^2}{m_H^2}\right)^2(q_f^2C_f)^2(4m_{\nu}^2-t)|{\cal
B}^{\pr}(t)|^2\,,
\end{equation}
which leads to the differential cross section
\begin{equation}\label{hdsdt}
\frac{d\sigma}{dt}=\frac{G_F^2m_{\nu}^2\al^2}{\pi^3}\left(\frac{m_f^2}{m_H^2}\right)^2(q_f^2C_f)^2\frac{(4m_{\nu}^2-t)}{(s-m_{\nu}^2)^2}|{\cal
B}^{\pr}(t)|^2\,.
\end{equation}

Before evaluating Eq.\,(\ref{hdsdt}) in detail, recall that the
fermion-Higgs coupling is proportional to the fermion mass $m_f$,
in which case the contribution from the heaviest quark dominates.
Since $|t|\leq (s-m_{\nu}^2)^2/s\sim 1.0\,$ MeV, we have
$-t<<m_f^2$, and
\begin{equation}
{\cal B}^{\pr}(t)\to \frac{1}{6}\frac{t}{m_f^2}\,.
\end{equation}
This implies that the leading contribution to the total cross
section from $H^0$ exchange behaves as
\begin{equation}
\sigma\sim
\frac{G_F^2m_{\nu}^2\al^2}{\pi^3}\frac{(s-m_{\nu}^2)^6}{m_H^4s^4},
\end{equation}
which is a factor $(m_e^2/m_H^2)^2$ smaller than the leading order
term from $Z^0$ exchange, Eq.\,(\ref{siglead}), and hence
completely negligible. This analysis shows that the cross section
is dominated by the diagram with the lowest mass particle, the
electron, in the loop and even this contribution is further
suppressed by the anomaly cancellation mechanism. In the same way,
the triangle diagrams with $W$'s in the loop are also negligible.
The remainder of our discussion is based on using the cross
section Eq.\,(\ref{sigtot}).

\section{Neutrino Factories and the process  $\bm{\gamma\nu\to\gamma\nu}$ }

A neutrino factory, such as that described in the Neutrino Factory and
Muon Collaboration (NFMC) feasibility report\cite{nfmc}, may be the only
viable way to study neutrino-photon scattering.  We use the machine
design described in that report as our general guideline for estimating
the event rate.  Unfortunately, the cross section we have obtained is
discouragingly small and major improvements in laser and
storage ring technology will be required in order to make this process
accessible at a future neutrino factory.   Our aim is to delineate the
basic requirements for the study of this process.  We hope that our
somewhat naive theoretical projections will motivate a more careful and
experimentally more realistic study.

We take as our baseline equipment a muon storage ring such as that
described in the NFMC report\cite{nfmc}.  Such a machine would be a
first step in the construction and operation of a future muon
collider.  One of the designs mentioned in the report has a race track
shaped storage ring with a total circumference of 1 km.  Each of the
straight sections has a length of about one quarter the circumference.
The muon energy would be 50 GeV and the muon flux is projected to be
of the order of a millimole per year.  A highly energetic, high flux
neutrino beam will be generated by the decay of $10^{20}-10^{21}$
muons.  The neutrinos generated along the straight sections will be
highly collimated.  The actual size of the storage ring is a function
of the muon beam energy, in order to consider a variety of beam
energies we will make use of the expression given in
reference~\cite{geer} for the storage ring circumference in meters,
\begin{equation}\label{eq:L}
  L\simeq \frac{60E_\mu}{B},
\end{equation}
where $B$ is the magnetic field for the bending magnets in units of Tesla
and $E_\mu$ is the muon energy in units of GeV.  For the photon source we
envision some type of high powered laser which would be placed close to
the ring and aimed directly at the muon/neutrino beam along one of the
linear segments of the ring.  The analysis presented below is grounded on
this basic experimental set-up. It is possible that a more ingenious
set-up could improve the prospects for observing neutrino-photon
interactions.

The rate of neutrino-photon scatterings can be expressed as,
\begin{equation}
        R = f_R N_\nu n_\gamma \sigma,
\end{equation}
where $N_\nu$ is the number of neutrinos per bunch which overlaps with
the photon beam (Eq.~(\ref{eq:nnu})), $n_\gamma$ is the number
of photons per cross sectional area, $f_R$ is the repetition rate or
number of bunches per second, and $\sigma$ is the cross section
(Eq.~(\ref{sigtot})).  For simplicity we will assume that the
repetition rate for the laser, like that of the muon storage ring, is
equal to 15 Hz.

The properties of the neutrino beam are well defined by the muon
energy $E_\mu$.   In the muon rest frame the maximum muon neutrino energy
is given by, $E^*_{\nu_\mu}=m_\mu/2$, the energy distribution peaks at this
maximum value, and has an average value which is seventy percent
of the maximum value. In the lab frame the neutrino energy is then,
$E^{lab}_{\nu_\mu}\simeq E_\mu/3$.   Polarization effects in this
process are small and we will ignore them in what follows.

In the lab frame the polar angle, measured with respect to the beam
direction, is related to the CM polar angle by,
\begin{equation}
 \theta_l \simeq \frac{m_{\mu} \,\tan(\theta_{\rm cm}/2)}{E_\mu},
\end{equation}
where on average $\tan(\theta_{\rm cm}/2)$ is close to 1.0.  We assume
that this is the dominant source of divergence in the neutrino beam. The
width of the neutrino beam at the interaction region is then of the order
of,
\begin{equation}
d_{\nu}\sim \frac{z m_{\mu}}{E_{\mu}}\,
\end{equation}
where  $z$ is the
distance from the point of decay to the interaction region.

Given a fixed area $A$, determined by the width of the photon beam,
only a fraction of the muon decays along the straight section will
contribute to the scattering process.  Decays which are
closer to the interaction region will generate a larger fraction of
neutrinos which could lead to neutrino photon scattering than decays
which are further away.  In order to determine that fraction, $\eta$, we
will assume that all decays at a given decay point generate a gaussian
shaped neutrino beam whose width is of the order of $d_{\nu}$.  Then
the fraction of neutrinos which originate from muon decays at a distance
$z$ from the interaction region, and which fall within a distance
$\delta\simeq \sqrt{A}$ of the center of the beam,
is given by,
\begin{equation}
P(z,\delta)=\frac{2}{\sqrt{\pi}}\int_0^{\delta/\sqrt{2}d_\nu}du\,e^{-u^2}\,.
\end{equation}
Furthermore, the fraction of all muon decays that occur over a length $\ell$ and
which give rise
to muon neutrinos  that fall within a distance $\delta$ of the center
of the beam is then,
\begin{equation}
\eta = \frac{1}{\ell{}^\prime}\int_0^{\ell\,'}dz\,P(z,\delta),
\end{equation}
where, $\ell'=\ell+D$, and $D$ is the distance from the end of the
straight section to the interaction region.   We will assume that $D$ is
of the order of 30 m or less \cite{mcfarlane}. For the energies considered
here the effects of $D$ are small, smaller than 25\% for $E_{\mu}=30$ GeV.
The effects become less important at higher energies and we ignore $D$
from this point on. We also assume that  the length of the straight
section is fixed at 25\% of the total circumference\cite{geer}. Therefore,
if there are $N_{B}$ muons per bunch, and we assume that the muons are
equally likely to decay anywhere along the ring, the number of neutrinos
falling within a distance $\delta$ of the center of the beam is given by,
\begin{equation}
\label{eq:nnu}
N_\nu=\frac{1}{4}\,\eta N_{B}.
\end{equation}
For our estimates, we use $N_B=10^{12}$\,\cite{quigg}.

With these assumptions and the use of Eq.\,(\ref{eq:L}), we can see that
$\eta$ is roughly independent of the muon energy.  The expression for
$\eta$ can be rewritten in terms of dimensionless integral as,
\begin{equation}
 \eta = \xi\int_0^\frac{1}{\xi}dx\,\erf(1/x)\,,
\end{equation}
where $\erf(x)$ is the error function associated with the normal
distribution and the dimensionless parameter $\xi$  is defined by,
\begin{equation}
 \xi = \frac{2\sqrt{2}\,\delta E_\mu}{m_\mu\, L}.
\end{equation}
Using Eq.~(\ref{eq:L}) gives
\begin{equation}\label{delt}
 \xi = .45\;\delta B\,,
\end{equation}
with $\delta$ in meters and $B$ in Tesla. Thus, with the assumptions used
here, $\eta$ is a function of $\delta$ and $B$, i.e.
$\eta=\eta(\delta,B)$.

\begin{figure}
\hfill\includegraphics[height=2.4in]{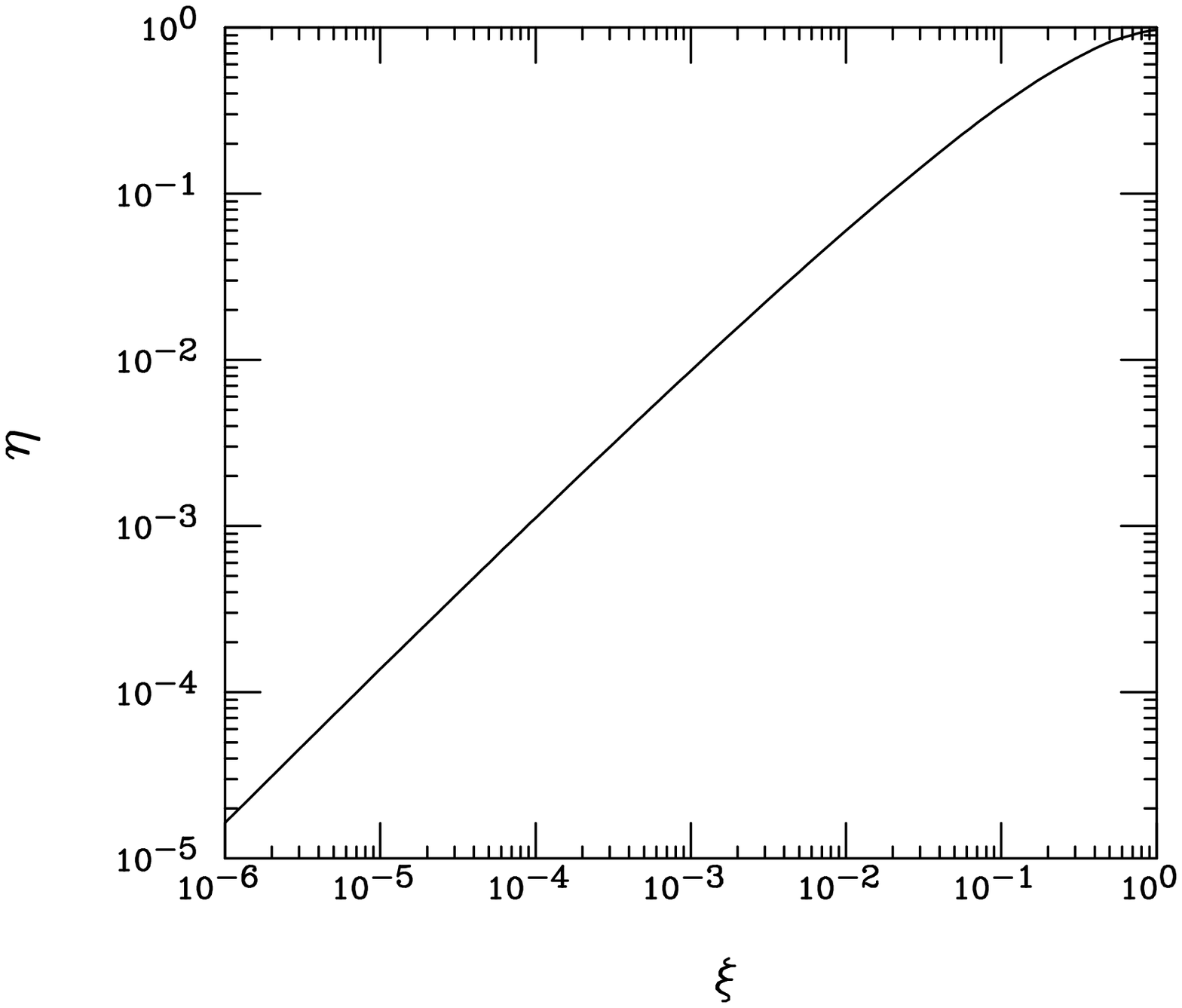}%
\hfill\includegraphics[height=2.4in]{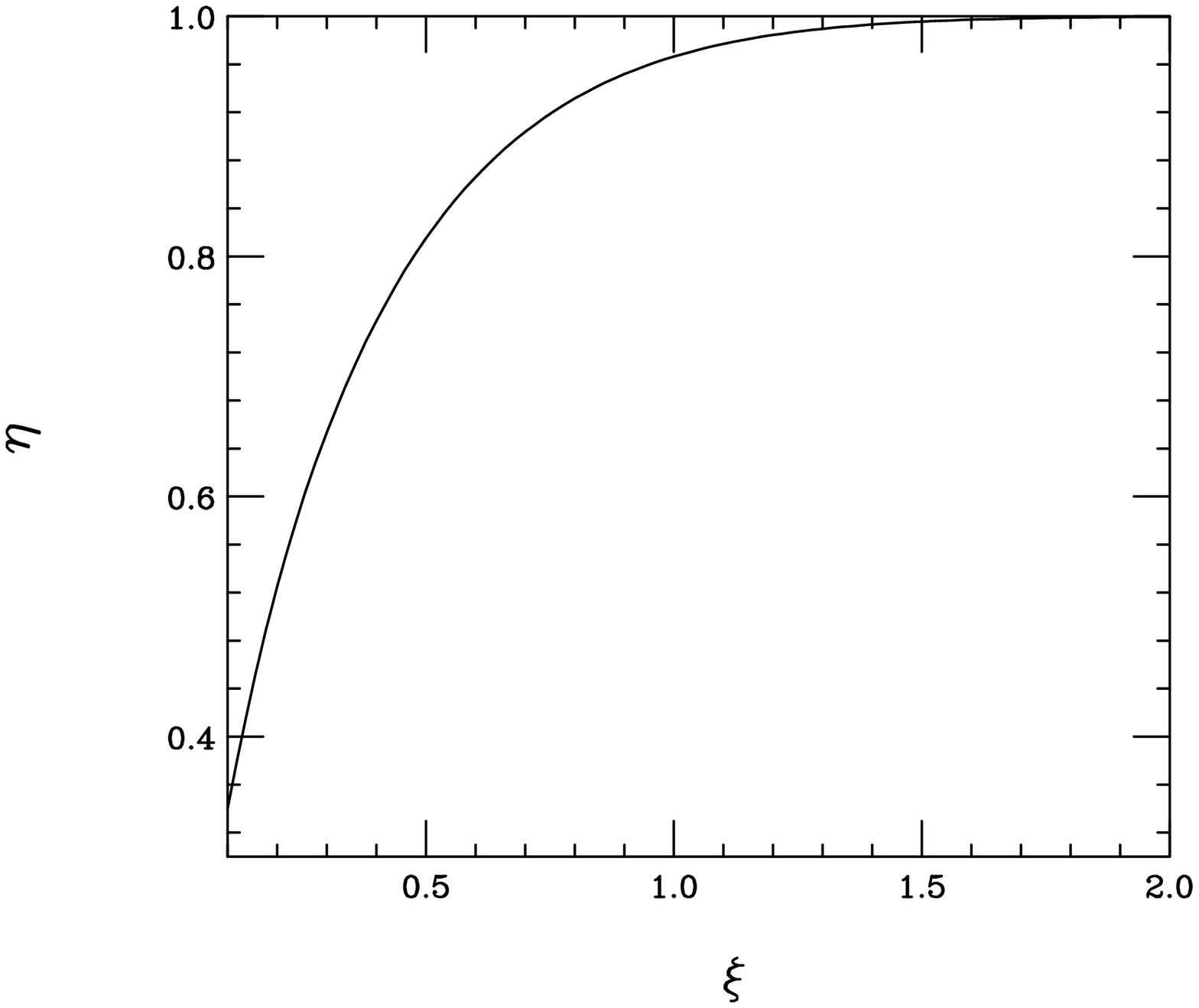}\hspace*{0pt\hfill}
\caption{The fraction $\eta$ of muon neutrinos which will
  contribute to neutrino-photon scattering within a circle of radius
  $\delta$ centered on the center of the neutrino beam is shown. The variable
  $\xi$ is related to $\delta$ by Eq.~(\ref{delt}) in the text.  The
  right panel is a magnified view for the range $.1<\xi<2$.
\label{fractn}}
\end{figure}

In Fig.~\ref{fractn} we present a plot of $\eta$ for a range of values
of $\xi$.  From this figure and the definition of $\xi$ one can
determine the fraction $\eta$ for any given set of storage ring
parameters. For example from the figure we can determine that, given a
bending magnetic field of 8 T (3 T), sixty percent of the decays
within the straight section will fall within a circle of radius 7 cm
(18 cm).  Therefore, one way to increase $\eta$ and thus improve the
results presented below is to increase $B$ or equivalently have the
ratio $L/E_{\mu}$ be as small as possible.

In regard to the laser system we have used the parameters for
ultra-powerful plasma lasers described in the work of Malkin, Shvets, and
Fisch~\cite{malkin}.  In this speculative work Malkin {\em et al.} report
that short pulsed lasers with energies of up to $1.6\times 10^7$ J per
pulse may be possible.  These high energies require pulsed-lasers with
very short pulse durations ($\sim 1$ fs). In principle, these lasers can
reach densities up to the critical value of $10^{34}$
eV/cm$^2$\cite{malkin}. At such high densities the electromagnetic fields
are large enough to produce electron-positron pairs and induce vaccuum
breakdown.  This number represents an absolute limit for lasers and thus
imposes a strong restriction on how well we will be able to use lasers to
study neutrino-photon interactions.

Our results are summarized in Tables~\ref{tab:3} and~\ref{tab:4}.  The
results were obtained assuming an energy per pulse of $1.6\times 10^7$ J
for the laser, a muon neutrino mass of $100$ keV, and a bending magnetic
field of 8 T.  In Table~\ref{tab:3} we use, $E_\gamma=2.6$ eV, and present
the results for several neutrino energies and two representative choices
of the beam width.  In the first case we set the beam width equal to five
times the photon wavelength, $\delta=5\lambda$.  In this case
$\eta=1.2\times 10^{-4}$ and the energy density is $1.8\times 10^{33}$
eV/cm$^2$. For the second case we use $\delta=1$ cm; the fraction is then
$\eta=1.6\times 10^{-1}$ and the energy density is $10^{26}$ eV/cm$^2$. To
obtain one event per year the rate should be about $3\times 10^{-8}$
s$^{-1}$. From the fourth and fifth columns we can see that even with the
optimistic laser parameters we have used here the results are five to six
orders of magnitude too small for observability.  Note that in both cases
the photon energy densities are close to the critical limit so there is
not much room for improvement on the laser side.  Note also that
increasing the beam size so as to increase the fraction $\eta$ only makes
things worse. Furthermore, increasing the neutrino energy helps very
little; this remains true as long as Eq.~\ref{eq:L} remains valid.  The
best improvement would come from increasing the neutrino flux.

In Table~\ref{tab:4} we present the results with the higher photon energy
proposed in reference~\cite{malkin}, $E_\gamma=50$ eV.  Again the results
for several neutrino energies and two representative choices of the beam
width are presented.  For the first case we follow Ref.\,\cite{malkin} and
set the beam width equal to forty times the photon wavelength,
$\delta=40\lambda$.  In this case $\eta=5.3\times 10^{-5}$ and the energy
density is at the critical limit of $10^{34}$ eV/cm$^2$.  For the second
case we use the same parameters as in the second case of
Table~\ref{tab:3}.  Results are of the same order of magnitude as those in
Table~\ref{tab:3}.

\section{Discussion}

The cross section for neutrino-photon scattering with massive neutrinos is
greatly enhanced over the standard model cross section for massless
neutrinos. Using the cross section for massive neutrinos, which scales as
$m_{\nu}^2$, we have explored the feasibility of measuring the muon
neutrino mass in at a future neutrino factory.

The results for the reaction rate are five to six orders of magnitude too
small to be observable at the rate of 1 event/year. This result was
obtained despite having assumed rather optimistic parameters for the laser
system. Currently the highest energy laser, the pentawatt laser at
LBL\cite{mourou}, has an energy per pulse of $10^3$ J and a repetition
rate of only .008 Hz.  However, lasers with peak powers of the order of
exawatts and high repetition rates ($\sim 10$ Hz) are within reach of
current technologies\cite{mourou}.  Therefore, the laser systems described
in ref.~\cite{malkin} may not be far off.

Assuming these great advances in laser technology, it would still be
necessary to improve the neutrino flux by five to six orders of magnitude
in order to study neutrino photon scattering using the approach suggested
here.  Further improvements may be obtained by decreasing the $L/E_{\mu}$
ratio or increasing the bending magnetic field $B$ from the 8 T assumed
here.   Finally, more complicated arrangements with several lasers located
along the circumference storage beam may serve to overcome the
deficiencies of our simpler approach.

Observation of the scattering event is relatively easy because
neutrino-photon scattering results predominantly in a back scattered high
energy photon (see Fig.\,\ref{dist}) and the standard photon detection
techniques, e.g. crystal calorimetry, can be used with high efficiency and
low background\cite{rich}.

\begin{acknowledgements}
We would like to thank A.C. Melissinos whose initial questions on neutrino
photon scattering motivated this work.  We are grateful for the full
participation and cooperation by R. Stroynowski and V.  Teplitz throughout
most of this work. S. Rinaldi and J. Rothenberg provided guidance on the
latest developments in lasers. This research is funded in part by the
National Science Foundation under grant PHY-9802439 and by the Department
of Energy under contracts DE-FG03-95ER40908 and DE-FG03-93ER40757.
\end{acknowledgements}

\begin{center}
\section*{Tables}
\end{center}
\begin{table}[h]
\begin{center}
\begin{tabular}{c@{\hspace{10pt}}c@{\hspace{10pt}}c}  \toprule
 $E_{\nu}$ (GeV)   & $\sigma$ (fb)  &$\sigma_{\rm Leading}$ (fb) \\ \colrule
  5        & $3.797\times 10^{-17}$ & $4.422\times 10^{-17}$  \\ \colrule
 10        & $1.448\times 10^{-16}$ & $1.948\times 10^{-16}$  \\ \colrule
 25        & $6.674\times 10^{-16}$ & $1.291\times 10^{-15}$  \\ \colrule
 50        & $1.711\times 10^{-15}$ & $5.269\times 10^{-15}$  \\ \botrule
\end{tabular}
\end{center}
\vspace{-8pt} \caption{A comparison of the exact cross section and the leading
order approximation is given for several values of the neutrino energy
$E_{\nu}$ \label{tab:1}}
\end{table}

\begin{table}[h]
\begin{center}
\begin{tabular}{c@{\hspace{20pt}}c}  \toprule
 $E_{\nu}$ (GeV)   & $\omega$ (eV)   \\ \colrule
 10 &785 \\ \colrule
 15 &523 \\ \colrule
 20 &393 \\ \colrule
 50 &157 \\ \botrule
\end{tabular}
\end{center}
\vspace{-8pt} \caption{\label{tab:2} The laboratory photon energy $\omega$
at which the maximum cross section $\sigma_{\mathrm{max}}=8.89\times
10^{-54}\mathrm{cm^2}$ occurs is given as a function of the neutrino
energy $E_{\nu}$. }
\end{table}
%

\begin{table}[h]
\begin{center}
\begin{tabular}{c@{\hspace{12pt}}d@{\hspace{12pt}}c@{\hspace{12pt}}c@{\hspace{12pt}}c}
\toprule
\multicolumn{1}{c}{$E_{\nu}$ (GeV)} &
\multicolumn{1}{c}{$\sqrt{s}$ (MeV)} &
\multicolumn{1}{c}{$\sigma$ (cm$^2$)} & \multicolumn{2}{c}{$R$ (s$^{-1}$)} \\ 
\multicolumn{1}{c}{} &\multicolumn{1}{c}{}&\multicolumn{1}{c}{}&
\multicolumn{1}{c}{$\delta=5\lambda_\gamma$} &\multicolumn{1}{c}{$\delta=1$ cm}
\\ \colrule
$3.00\times 10^{1}$ & .567 &  $1.00\times 10^{-55}$ & $3.0\times 10^{-14}$ & $2.3\times 10^{-18}$ \\
\colrule
$5.00\times 10^{1}$ & .728 &  $2.50\times 10^{-55}$ & $7.6\times 10^{-14}$ & $5.8\times 10^{-18}$ \\
\colrule
$2.50\times 10^{2}$ & 1.62 &  $2.40\times 10^{-54}$ & $7.3\times 10^{-13}$ & $5.5\times 10^{-17}$ \\
\colrule
$1.00\times 10^{3}$ & 3.23 &  $7.10\times 10^{-54}$ & $2.2\times 10^{-12}$ & $1.6\times 10^{-16}$ \\
\colrule
$1.00\times 10^{5}$ & 32.2 &  $8.60\times 10^{-55}$ & $2.6\times 10^{-13}$ & $2.0\times 10^{-17}$ \\
\botrule
\end{tabular}
\end{center}
\vspace{-8pt} \caption{\label{tab:3} The reaction rate $R$ for
neutrino-photon scattering is shown. The photon energy is fixed at
$E_\gamma=2.6$ eV ($\lambda_\gamma=476.7$ nm), the neutrino mass is taken
to be 100 keV and the bending magnetic field is 8 T. For the laser, the
bunch energy is fixed at $1.6\times 10^7$ J, and a repetition rate of 15
Hz is assumed. Results are presented for several muon neutrino energies
and two distinct values for $\delta$: $\delta=5\lambda_\gamma$ and
$\delta=1$ cm. }
\end{table}
\begin{table}
\begin{center}
\begin{tabular}{c@{\hspace{12pt}}d@{\hspace{12pt}}c@{\hspace{12pt}}c@{\hspace{12pt}}c}
\toprule
\multicolumn{1}{c}{$E_{\nu}$ (GeV)} &
\multicolumn{1}{c}{$\sqrt{s}$ (MeV)} &
\multicolumn{1}{c}{$\sigma$ (cm$^2$)} & \multicolumn{2}{c}{$R$ (s$^{-1}$)} \\ 
\multicolumn{1}{c}{} &\multicolumn{1}{c}{}&\multicolumn{1}{c}{}&
\multicolumn{1}{c}{$\delta=40\lambda_\gamma$} &\multicolumn{1}{c}{$\delta=1$ cm}
\\ \colrule
$3.00\times 10^{1}$ & 2.45 & $5.10\times 10^{-54}$ & $2.1\times 10^{-13}$ & $6.1\times 10^{-18}$ \\
\colrule
$5.00\times 10^{1}$ & 3.16 & $6.90\times 10^{-54}$ & $2.8\times 10^{-13}$ & $8.3\times 10^{-18}$ \\
\colrule
$1.50\times 10^{2}$ & 5.48 & $9.10\times 10^{-54}$ & $3.7\times 10^{-13}$ & $1.1\times 10^{-17}$ \\
\colrule
$2.50\times 10^{2}$ & 7.07 & $8.60\times 10^{-54}$ & $3.5\times 10^{-13}$ & $1.0\times 10^{-17}$ \\
\colrule
$1.00\times 10^{3}$ & 14.10 & $4.40\times 10^{-54}$ & $1.8\times 10^{-13}$ & $5.3\times 10^{-18}$ \\
\botrule
\end{tabular}
\end{center}
\vspace{-8pt} \caption{\label{tab:4} The reaction rate $R$ for
neutrino-photon scattering is shown.  The photon energy is fixed at
$E_\gamma=50$ eV ($\lambda_\gamma=24.79$ nm), and
$\delta=40\lambda_{\gamma}$. The other parameters are the same as in
Table~\ref{tab:3}.}
\end{table}


\begin{thebibliography}{99}
\bibitem{conrad} J.M. Conrad, Proceedings of the 29th International Conference on High Energy
Physics, Vancouver, Canada, July, 1998.
\bibitem{SLAC} Plenary talk at the XIX International Symposium on Lepton and Photon
Interactions, Stanford, California, August, 1999.  hep-ex 9912007.
\bibitem{pdg} Review of Particle Physics, The European Physical Journal {\bf
C3}, 1 (1998).
\bibitem{DR} D. A. Dicus and W. W. Repko, Phys. Rev. D {\bf 48}, 5106 (1993);
D. A. Dicus and W. W. Repko, Phys. Rev. Lett. {\bf 79}, 569 (1997); D. A.
Dicus, C. Kao and W. W. Repko, Phys. Rev. D {\bf 59}, 013005 (1999).
\bibitem{teplitz} M. Harris, J. Wang and V. Teplitz, astro-ph/9707113
(unpublished).
\bibitem{sha} R. Shaisultanov, Phys. Rev. Lett. {\bf 80}, 1586 (1998).
\bibitem{chyi} T.-K. Chyi {\em et al.}, hep-ph/9907384.
\bibitem{Yang} C. N. Yang, Phys. Rev. {\bf 77}, 242 (1950).
\bibitem{GellMann} M. Gell-Mann, Phys. Rev. Lett. {\bf 6}, 70 (1961).
\bibitem{nfmc} K.T. McDonald {\em et al.}, hep-ph/9911009.
\bibitem{mcfarlane} K. McFarland,
http://www.pas.rochester.edu/$^{\sim}$ksmcf/musr, Fermilab Neutrino
Factory Physics Study, February 18, 2000.
\bibitem{geer} S. Geer, Phys. Rev. D {\bf 57}, 6989 (1998).
\bibitem{quigg} C. Quigg, {\em Physics with a Millimole of Muons},
hep-ph/9803326.
\bibitem{malkin} V.M. Malkin, G. Shvets, and N.J. Fisch, Phys. Plasmas
{\bf 7}, 2232 (2000).
\bibitem{mourou} G.A. Mourou, C.P.J. Barty, and M.D. Perry, Phys. Today {\bf 51}, 22
(Jan. 1998).
\bibitem{rich} R. Stroynowski, Private conversation.
\end{thebibliography}
\end{document}